\documentclass[preprint,authoryear]{sigplanconf}

\usepackage{hyperref}
\usepackage{amsmath,amsthm}
\usepackage{hypernat}
\usepackage{xcolor}
\usepackage[utf8]{inputenc}
\usepackage{listings}

\usepackage{tikz}
\usepackage[skins,listings]{tcolorbox}

\tcbset{listing engine=listings}
\tcbset{listingbox/.style={%
    enhanced,
    boxsep=-9pt,
    left=15pt,
    arc=2pt,
    boxrule=1pt,
  }
}

\definecolor{green}{RGB}{0, 180, 0}

\lstdefinestyle{custompython}{%
    belowcaptionskip=1\baselineskip,
    breaklines=true,
    frame=none,
    xleftmargin=\parindent,
    language=Python,
    showstringspaces=false,
    basicstyle=\footnotesize\ttfamily,
    keywordstyle=\bfseries\color{green!40!black},
    commentstyle=\itshape\color{purple},
    identifierstyle=\color{black},
    stringstyle=\color{green!60!black},
    keywords=[2]{as,True,False},
    keywordstyle=[2]\bfseries\color{green!40!black},
    keywords=[3]{np,sp,plt},
    keywordstyle=[3]\bfseries\color{blue},
    numbers=none,
    columns=fullflexible,
}
\lstdefinestyle{customfortran}{%
    belowcaptionskip=1\baselineskip,
    breaklines=true,
    frame=none,
    xleftmargin=\parindent,
    language=Fortran,
    showstringspaces=false,
    basicstyle=\footnotesize\ttfamily,
    keywordstyle=\bfseries\color{green!60!black},
    commentstyle=\itshape\color{purple},
    identifierstyle=\color{black},
    stringstyle=\color{green},
    numbers=none,
    columns=flexible,
}
\lstdefinestyle{customc}{%
    belowcaptionskip=1\baselineskip,
    breaklines=true,
    frame=none,
    xleftmargin=\parindent,
    language=C,
    showstringspaces=false,
    basicstyle=\footnotesize\ttfamily,
    keywordstyle=\bfseries\color{green!60!black},
    commentstyle=\itshape\color{purple},
    identifierstyle=\color{black},
    stringstyle=\color{green},
    numbers=none,
    columns=fullflexible,
}

\newtcblisting{mylisting}{%
  listingbox,
  listing only,
  listing options={style=custompython},
}

\def\loopy{Loo.py}

\begin{document}


\setlength{\pdfpageheight}{\paperheight}
\setlength{\pdfpagewidth}{\paperwidth}

\conferenceinfo{ARRAY'15}{June 13, 2015, Portland, OR, USA}
\CopyrightYear{2015}
\crdata{978-1-4503-3584-3/15/06}
\doi{10.1145/2774959.2774969}


\exclusivelicense%


\titlebanner{\loopy: From Fortran to performance}        
\preprintfooter{Submitted to \href{http://www.sable.mcgill.ca/array/}{ARRAY15}}   

\title{%
\loopy: From Fortran to performance via transformation and substitution rules}

\authorinfo{Andreas Klöckner}
           {University of Illinois at Urbana-Champaign}
           {andreask@illinois.edu}

\maketitle

\begin{abstract}
  A large amount of numerically-oriented code is written and is
  \emph{being} written in legacy languages. Much of this code could, in principle,
  make good use of data-parallel throughput-oriented computer
  architectures. \loopy, a transformation-based programming system targeted
  at GPUs and general data-parallel architectures, provides a mechanism for
  user-controlled transformation of array programs. This transformation
  capability is designed to not just apply to programs written specifically
  for \loopy, but also those imported from other languages such as
  Fortran\@.
  It eases the trade-off between achieving high performance, portability, and
  programmability by allowing the user to apply a large and growing family
  of transformations to an input program. These transformations are expressed
  in and used from Python and may be applied from a variety of settings,
  including a pragma-like manner from other languages.
\end{abstract}

\category{D}{3}{4} --- Code generators
\category{D}{1}{3} --- Concurrent programming
\category{G}{4}{} --- Mathematical software




\section{Introduction}
\label{sec:intro}
\loopy\ \citep{kloeckner_loopy_2014} is a programming system for array
computations that targets CPUs, GPUs, and other, potentially heterogeneous
compute architectures. One salient feature of \loopy\ is that programs
written in it necessarily consist of two parts:
\begin{itemize}
\item A semi-mathematical statement of the array computation to be carried
  out, in terms of a \emph{loop polyhedron} and a partially ordered \emph{set
  of `instructions'}.
\item A sequence of \emph{kernel transformations}, driven by an `outer'
  program in the high-level scripting language Python \citep{vanrossum_python_1994}.
\end{itemize}
This strong separation is an explicit design goal, as it enables specialization of
users, cleanliness of notation in either part, as well as greater
flexibility in terms of transformation.

While a prior article\citep{kloeckner_loopy_2014} emphasized \loopy's
program model and semantics, this article focuses on the
transformation-related aspects of the system.

\loopy\ was designed to suit a number of different use cases, all of which
have shaped its design:
\begin{itemize}
  \item a means to concisely express computational kernels
    in the design of scientific computing applications (such as
    solvers for partial differential equations
    \citep{kloeckner_nodal_2009}),
  \item a foundation for outlining the search space to be explored by
    an autotuning component or a human performance tuner,
  \item an on-the-fly code generator for computational software,
  \item a code-generation back-end enabling high-level DSLs
    to obtain performance on heterogeneous architectures, and
  \item a program transformation tool for de- and re-optimizing
    legacy code.
\end{itemize}
The present article demonstrates how \loopy\ can function as a code
generation back-end for a subset of Fortran (as an example of a language
separate from \loopy's own internal representation) while maintaining its
full capability to transform the ingested code in a manner comprehensible
and useful to the author of the original program. A number of mechanisms
are described that are intended to aid the formulation of transformations
on array computations in this setting.

As one example of the issues that arise, the strong separation of semantics
and transformation, while desirable, also poses a difficulty. For example,
unlike in an annotation-based setting, where lexical proximity alone can be
used to indicate what part of a program is to be transformed, this option
does not exist for \loopy, and so alternatives have to be devised.

The literature on code generation and optimization for array
languages is vast, and no attempt will be made to provide a survey of the
subject in any meaningful way. Instead, we will seek to highlight a few
approaches that have significantly influenced the thinking behind \loopy,
are particularly similar, or provide ideas for further development.
\loopy\ is heavily inspired by the polyhedral model of expressing
static-control programs \citep{feautrier_automatic_1996,bastoul_code_2004}.
While it takes significant inspiration from this approach, the details of
how a program is represented, beyond the existence of a loop domain, are
quite different.  High-performance compilation for GPUs, by now, is hardly
a new topic, and many different approaches have been used, including
ones using OpenMP-style directives
\citep{lee_openmpc_2010,han_hicuda_2011},
ones that are fully automatic \citep{yang_gpgpu_2010},
ones based on functional languages \citep{svensson_obsidian_2010}, and
ones based on the polyhedral model \citep{verdoolaege_polyhedral_2013}.
Other ones
define an automatic, array computation middleware
\citep{garg_velociraptor_2012} designed as a back-end for multiple languages,
including Python. Automatic, GPU-targeted compilers for languages embedded
in Python also abound
\citep{catanzaro_copperhead_2011,rubinsteyn_parakeet_2012,continuum_numba_2014},
most of which transform a Python AST at run-time based on various
levels of annotation and operational abstraction.

Code generators just targeting one or a few specific workloads (often
matrix-matrix multiplication) using many of the same techniques available
in \loopy\ have been presented by various authors, ranging from early work
such PhiPAC
\citep{bilmes_optimizing_1997} to more recent OpenCL- and CUDA-based work
\citep{cui_automatic_2011,matsumoto_implementing_2012}.

Other optimizing compilers assume a substantial amount of domain knowledge
(such as what is needed for assembly of finite element matrices) and
leverage this to obtain parallel, optimized code. One example of this
family of code generators is COFFEE \cite{luporini_crossloop_2015}.

Perhaps the conceptually closest prior work to the approach taken by
\loopy\ is CUDA-CHiLL \citep{rudy_programming_2011}, which performs
source-to-source translation based on a set of user-controlled
transformations \citep{chen_chill_2008,hall_loop_2010}. \loopy\ and CHiLL
still are not quite alike, using dissimilar intermediate representations,
dissimilar levels of abstraction in the description of transformations, and
a dissimilar (static vs.\ program-controlled) approach to transformation.

Source-to-source transformation similarly has been studied extensively,
with many mature systems existing in the literature
(for instance \citep{schordan_source_2003,dave_cetus_2009}).


\section{\loopy's view of a kernel}
\label{sec:loopy-kernel}
We begin by briefly examining \loopy's model of a program (or `kernel').  A
very simple
example of a kernel shall serve as an introduction.  This kernel
reads in one vector, doubles it, and writes the result to another:
\begin{tcolorbox}[listingbox]
\begin{lstlisting}[style=custompython,gobble=2]
  knl = loopy.make_kernel(
      "{[i]: 0<=i<n}",  # loop domain
      "out[i] = 2*a[i]")  # instructions
\end{lstlisting}
\end{tcolorbox}
\noindent
The above snippet of code illustrates the main components of a \loopy\
kernel:
\begin{itemize}
  \item The \emph{loop domain}: \verb|{ [i]: 0<=i<n }|. This defines
    the integer values of the loop variables for which instructions
    (see below) will be executed.
    It is written in the syntax of the \texttt{isl} library
    \citep{verdoolaege_isl_2010}.  \loopy\ calls the loop variables
    \emph{inames}. In this case, \verb|i| is the sole iname.
    \verb|n| is a \emph{parameter} that is passed to the kernel by the
    user. \verb|n| in this case determines the length of the vector being
    operated on.

    To accommodate some data-dependent control flow, there is not actually
    a single loop domain, but rather a \emph{tree of loop domains},
    allowing more deeply nested domains to depend on inames
    introduced by domains closer to the root.

  \item The \emph{instructions} to be executed: \verb|out[i] = 2*a[i]|. These are scalar
  assignments between array elements, consisting of a left-hand
  side assignee and a right-hand side expression.
  Right-hand side expressions are allowed to contain the usual mathematical
  operators, calls to externally defined functions, and references to
  substitution rules (see Section~\ref{sec:subst-rules}).

  In addition to the left-hand- and right-hand-side expressions describing
  the assignment, each instruction carries the following data:
  \begin{itemize}
    \item An \emph{instruction identifier}. A string that uniquely
    identifies each instruction. Automatically generated if not
    specified. In addition to specifying the entire unique ID, a `prefix'
    may also be specified, based on which a unique ID is generated.
    \item A set of \emph{instruction tags}.
    Used for transformation targeting (see Section~\ref{sec:ttp}).
    \item A \emph{set of inames} specifying within which loops
    this instruction is intended to be nested. A heuristic
    \citep{kloeckner_loopy_2014} is applied to automatically discover
    this information. The iname nesting may be overridden by the
    user if the heuristic does not yield the intended result.
    \item A set of \emph{instruction IDs depended upon}, i.e.\ required to be executed
    before the current instruction. As described in \citep{kloeckner_loopy_2014},
    these dependencies act at the innermost loop nesting level shared between
    the dependent and depended-upon instruction. Like the nest-within
    inames, a default set of dependencies is found by a heuristic that
    creates dependencies on instructions that write those variables that
    are read by this instruction.
    \item A set of \emph{predicates}, the conjunction of which determines
    the condition under which the instruction will be executed. Each
    predicate refers to a stored `flag' variable or its negation.
    This flag variable must have been set previously, and it serves as
    a source for automatically generated dependencies.
  \end{itemize}
\end{itemize}
\section{Transforming Fortran into \loopy}
\label{sec:fortran}
While \loopy's native intermediate representation is sufficiently abstract
and convenient that it is suited to being used directly by a
user/programmer, one main use case for \loopy\ is to be a back-end to other
systems whose result is a machine representation of an array computation.

To illustrate this use case, a
Fortran~\citep{backus_fortran_1957} front-end for \loopy\ is described in
the following. Along the way,
this front-end provides a convenient case study of what can be done
to enable program transformation in a setting where the structure of an
input program is not designed to be convenient for rather but rather
given by outside constraints, such as a decades-old standards document.

Based on a number
of restrictions (see below), the main objective of \loopy's Fortran
front-end is not (and cannot be) to be a fully standards-conforming Fortran
compiler. Instead, it seeks to lessen the burden of capturing \loopy\
kernels in \loopy's native representation, by providing an alternate input
format with which a user may be more familiar. The continuing dominance
of Fortran in scientific and engineering fields where computation is
applied further means that providing transformation avenues to
modern architectures is a possibly impactful way to leverage
these legacy codes on modern-day architectures.

The Fortran~77 model of computation is a surprisingly good match for
\loopy's input language, sharing not just the array-based view of the data
being operated on, but also much of its type system and its model of the
subroutine as the main unit of program functionality.

Compared to the more comprehensive Fortran~90, a number of restrictions
exist:
\begin{itemize}
  \item No early exits (\verb|EXIT|, \verb|CYCLE|, \verb|RETURN|),
    no mid-subroutine entry points (\verb|ENTRY|),
    limited data-dependent control flow \emph{(essential)}
  \item No guaranteed order between trips through a loop \emph{(essential)}
  \item Translation acts on a single subroutine, which will be translated
    to a single OpenCL compute kernel. \emph{(liftable)}
  \item No I/O, no calls to other subroutines \emph{(liftable)}
  \item No pointers, limited support for structured types \emph{(liftable)}
  \item No support for \verb|SAVE| and \verb|COMMON| data \emph{(liftable)}
  \item No array-level assignments and intrinsics \emph{(liftable)}
  \item No dynamic memory management \emph{(liftable)}
\end{itemize}
Each of the above restrictions is qualified with whether it is essential
and unlikely to be lifted in future revisions \emph{(essential)} or
a matter of further software development \emph{(liftable)}. Some of these
are a direct result of underlying limitations in the OpenCL kernel language
for which \loopy\ generates code.

The following example shows a Fortran kernel being  translated
by \loopy:
\begin{tcolorbox}[listingbox]
\begin{lstlisting}[style=customfortran,gobble=2]
  subroutine fill(out, a, n)
    implicit none
    real*8 a, out(n)
    integer n

    do i = 1, n
      out(i) = a
    end do
  end
  !$loopy begin transform
  ! fill = lp.split_iname(fill, "i", 128,
  !     outer_tag="g.0", inner_tag="l.0")
  !$loopy end transform
\end{lstlisting}
\end{tcolorbox}
\noindent
The code shows a straightforward vector fill kernel. Mainly the section between the
\verb|$loopy begin/end transform|
markers (in Fortran comments) is of note. This section consists of
\emph{Python} code, and the
Fortran subroutine defined above becomes available here as a \loopy\ kernel
object, under a Python identifier of the same name. At this point, the user is
free to use the entire transform vocabulary defined in \loopy\ (see
\citep{kloeckner_loopy_2014} and
Section~\ref{sec:trans-lang}) on their kernel.

The availability of all of the Python programming language for program
transformation sets \loopy\ apart from other `pragma'-type approaches to
annotation such as OpenMP \cite{dagum_openmp_1998} or OpenACC \cite{openacc_2011} as well
as from other transformation script approaches such as CHiLL
\cite{rudy_programming_2011,chen_chill_2008,hall_loop_2010}. The following
usage patterns are enabled by it:
\begin{itemize}
  \item \emph{Abstraction.} Users are enabled to build their own,
    higher-level, compound transformations that may be shared among a
    family of kernels. For instance, a number of transformations changing
    the data layout of a computation could (and, likely, should) be shared
    among a group of kernels accessing said data.
  \item \emph{Dynamism.} Being based on a full-featured programming
    language allows the transform code to respond to its environment in
    interesting, non-trivial ways. As a simple example, a different
    transform path may be chosen depending on the target device for which
    code is to be generated. Alternatively, the transform code may consult
    a performance model or a database regarding the most promising
    transforms to apply. It could also be part of an auto-tuning scheme.
  \item \emph{Introspection.} Transforms (and the code calling them) are at
    liberty to inspect and reason about the kernel code. For example, it is
    straightforward to write a loop over a set of variables being written
    in a certain code region and apply prefetching or a data layout
    transformation to them.
    This helps keep the transform code general, adaptable, and reusable.
\end{itemize}
These points emphasize the fact that \loopy\ can be employed as a
lower-level infrastructure component, providing enough expressive power
for higher-level, more abstract transformations built on top of it.

\loopy\ takes the following steps when translating a Fortran kernel:
\begin{itemize}
  \item When a \verb|do| loop is encountered, a new axis is added to the
    current loop domain. If necessary, the loop variable (`iname' in
    \loopy-speak) and all its uses will be renamed to ensure uniqueness.
  \item Fortran's scalar assignments and data type/dimension declarations
    map directly onto the corresponding features in \loopy.
  \item When an \verb|if|/\verb|then| block is encountered, a
    \emph{predicate} variable is created based on the condition in the
    \verb|if| statement, and all \loopy\ instructions created from the body
    of the \verb|if| block have the predicate variable (or its negation,
    for the \verb|else| sub-block) applied to it.
  \item Since Fortran programs are strongly sequentially ordered, the
    translation creates a linear chain of dependencies matching the program
    order.
\end{itemize}
A somewhat more challenging example including conditionals is shown below:
\begin{tcolorbox}[listingbox]
\begin{lstlisting}[style=customfortran,gobble=2]
  do i = 1, n
    a = inp(i)
    if (a.ge.3) then
        b = 2*a
        do j = 1,3
            b = 3 * b
        end do
        out(i) = 5*b
    else
        out(i) = 4*a
    endif
  end do
\end{lstlisting}
\end{tcolorbox}
\noindent
\loopy translates this to the following C/OpenCL kernel code:
\begin{tcolorbox}[listingbox]
\begin{lstlisting}[style=customc,gobble=2]
  for (int i = 0; i <= -1 + n; ++i)
  {
    a = inp[i];
    loopy_cond0 = a >= 3;
    if (loopy_cond0)
    {
      b = 2.0 * a;
      for (int j = 0; j <= 2; ++j)
        b = 3.0 * b;
      out[i] = 5.0 * b;
    }
    if (!loopy_cond0)
      out[i] = 4.0 * a;
  }
\end{lstlisting}
\end{tcolorbox}
\noindent
It is worth noting that instead of evaluating the conditional for each
instruction separately, \loopy's code generation stage is capable of
grouping, also across loop entries/exits, to help reduce the cost of
conditional execution.

\loopy\ instructions generated from segments of a Fortran program may have
\emph{tags} applied to them to ease their identification in the
transformation process. This interacts with the transformation
facilities in \loopy\ and allows them to be applied to subsets of the
program.
This is accomplished through the
\verb|!$loopy begin/end tagged| marker in a Fortran comment:
\begin{tcolorbox}[listingbox]
\begin{lstlisting}[style=customfortran,gobble=2]
  !$loopy begin tagged: input
  a = cos(alpha)*inp1(i) + sin(alpha)*inp2(i)
  b = -sin(alpha)*inp1(i) + cos(alpha)*inp2(i)
  !$loopy end tagged: input

  r = sqrt(a**2 + b**2)
  a = a/r
  b = b/r

  out1(i) = a
  out2(i) = b
\end{lstlisting}
\end{tcolorbox}
\noindent
This subset of the program can then be selected for transformation using
the \emph{match expression} `\verb|*$input|' See Section~\ref{sec:ttp} for
details.

\section{Transforming Array Computations}
\label{sec:trans-lang}
\loopy\ employs a number of strategies to allow the creation of
maintainable, logical, and readable transformation code. One important
aspect of this is \emph{transform targeting}, and a mechanism for
performing this function is discussed next.
\subsection{Substitution rules}
\label{sec:subst-rules}
\textbf{Semantics.}
In addition to instructions (see Section~\ref{sec:loopy-kernel}), \loopy\
kernels may contain `substitution rules', which, as their most basic
function, permit common subexpressions to be factored out and defined once.
In addition to simple subexpressions, substitution rules also support
parameters. The behavior of \loopy's substitution rule system is similar to
other macro systems, albeit no flow control is provided for use during expansion. A similarity
exists with the C preprocessor, although substitution rule processing takes
place at the level of the expression tree rather than the token stream.

Unless otherwise removed, substitution rules are automatically expanded
immediately before code generation. The following simple example
illustrates their use:
\begin{tcolorbox}[listingbox]
\begin{lstlisting}[style=custompython,gobble=2]
  lp.make_kernel(
    "{[i,j,n,n2]: 0<=i,j<npart and 0<=n,n2<3}",
    """
    grav_force(m, M, r) := -66.742*m*M/r**2

    <> radc = sqrt(sum(n, (x[i,n]-center[n])**2))
    <> rad_j = sqrt(sum(n2, (x[i,n2]-x[j,n2])**2))

    force[i] = grav_force(mass[i], massc, radc) + \
        sum(j, grav_force(mass[i], mass[j], rad_j))
    """)
\end{lstlisting}
\end{tcolorbox}
\noindent
In \loopy's native kernel language, substitution rules are differentiated
from assignment instructions by the use of a different assignment operator
(`\verb|:=|') and, optionally, the use of round parentheses on the
left-hand side of the assignment to delimit argument names.

In addition to providing a convenience for coding complex computations,
one major role of substitution rules in \loopy\ is to provide an additional
facility for attaching identifiers to parts of the computation.

\textbf{Creation.} While \loopy's built-in language includes facilities for
writing substitution rules directly, it is not reasonable to expect that
every programming system to which \loopy\ may be coupled will offer this
possibility---the Fortran front-end of Section~\ref{sec:fortran}
is one such example. To retain the specificity contributed by substitution
towards the transformation targeting problem (see
Section~\ref{sec:ttp}), \loopy\ provides several ways
of creating substitution rules from `bare code':
\begin{itemize}
  \item \emph{Unification.} Provided with a unification pattern,
    \loopy\ can locate all subexpressions unifiable with it and convert
    them to invocations of a newly-created substitution rule. For example,
    the two subexpressions involving \verb|b| in the assignment
    \begin{tcolorbox}[listingbox]
    \begin{lstlisting}[style=custompython,gobble=6]
      a[i] = 23*b[i]**2 + 25*b[i]**2
    \end{lstlisting}
    \end{tcolorbox}
    are unified by
    \begin{tcolorbox}[listingbox]
    \begin{lstlisting}[style=custompython,gobble=6]
      knl = lp.extract_subst(knl,
       "bsquare", "alpha*b[i]**2", parameters=("alpha",))
    \end{lstlisting}
    \end{tcolorbox}
    which rewrites them to
    \begin{tcolorbox}[listingbox]
    \begin{lstlisting}[style=custompython,gobble=6]
      bsquare(alpha) := alpha*b[i_0]**2
      a[i] = bsquare(23) + bsquare(25)
    \end{lstlisting}
    \end{tcolorbox}

  \item \emph{Wrapping of variable read access.} A particular
    example of unification, and in fact the most common one.
    \loopy\ can wrap any
    reading access to an array or scalar variable in a substitution rule.
    Combined with precomputation (Section~\ref{sec:precomp}), this
    provides a mechanism for prefetching of off-chip variables.
  \item \emph{Conversion of an assignment to temporary.} Temporary
    variables are often used to hold intermediate results for reuse.
    \loopy\ provides a facility to convert such an assignment into
    a substitution rule. For example, the (Fortran) code
    \begin{tcolorbox}[listingbox]
    \begin{lstlisting}[style=customfortran,gobble=6]
      do i = 1, n
        a(i) = 6*inp(i)
      enddo
      do i = 1, n
        out(i) = 5*a(i)
      end do
    \end{lstlisting}
    \end{tcolorbox}
    can be rewritten to
    \begin{tcolorbox}[listingbox]
    \begin{lstlisting}[style=custompython,gobble=6]
      a_subst(i) := 6*inp[i]
      out[i_1] = 5*a_subst(i_1)
    \end{lstlisting}
    \end{tcolorbox}
    using the \verb|temporary_to_subst| transformation. As one example,
    this process of transitioning through a rule enables the programmer
    to change the granularity or a precomputation to comprise a larger or
    smaller footprint of the iteration domain. In some sense, this
    undoes a common subexpression elimination and is thus a type of
    de-optimization.
\end{itemize}

\subsection{Transformation targeting}
\label{sec:ttp}
In transforming computational kernels, it is often undesirable to apply a
transformation to an entire kernel. Instead, the user may wish to express
specifically which instructions or which subexpressions a transform should
act upon. \loopy\ supports this use case by matching names/IDs and `tags' of
instructions and substitution rule invocations.

An example may help clarify this:
\begin{tcolorbox}[listingbox]
\begin{lstlisting}[style=custompython,gobble=2]
  f(x) := x*a[x]
  g(x) := 12 + f(x)
  h(x) := 1 + g(x) + 20*g$three(x)

  a[i] = h$one(i) * h$two(i)
\end{lstlisting}
\end{tcolorbox}
\noindent
Three (nested) substitution rules are defined, \verb|f|, \verb|g|, and \verb|h|.
Many of the substitution rule invocations have a `tag' applied to them (suffixed
onto the rule identifier with a dollar sign, e.g. `\verb|h$two|'). If
necessary, this makes each rule invocation
individually selectable. These tags have no influence on the meaning
of the program. They only serve to make locations in the code identifiable.

We apply the \verb|expand_subst| transformation (which simply expands a substitution rule)
to the
invocation of \verb|g| tagged \verb|three| \emph{within} the invocation of
\verb|h| tagged \verb|two|:
\begin{tcolorbox}[listingbox]
\begin{lstlisting}[style=custompython,gobble=2]
  knl = expand_subst(knl, "g$three < h$two")
\end{lstlisting}
\end{tcolorbox}
\noindent
More generally, a user may match arbitrary portions of the rule expansion
stack. The first component in the stack match expression (`\verb|g$three|'
in the above example) is necessarily the innermost level of expansion, and
outer levels are separated by the \verb|<| symbol. Each level consists of
the `main identifier', matching a substitution rule name or an instruction
ID, and the `tag', matching either an invocation tag on a substitution
rule, or an instruction tag. Each of the two parts also supports
shell-style wildcards. Multiple levels may be matched by an ellipsis
(\verb|innermost < ... < outer|).

The above example results in the following code:
\begin{tcolorbox}[listingbox]
\begin{lstlisting}[style=custompython,gobble=2]
  f(x) := x*a[x]
  g(x) := 12 + f(x)
  h(x) := 1 + g(x) + 20*g$three(x)
  h_0(x) := 1 + g(x) + 20*(12 + f(i))
  a[i] = h$one(i)*h_0$two(i)
\end{lstlisting}
\end{tcolorbox}
When expanding the specified invocation of \verb|g|, not all invocations of
\verb|h| (which contained the invocation of \verb|g|) were affected. As a
result, a new, separate version of \verb|h|, named \verb|h_0| was created,
and the relevant invocation sites of \verb|h| were updated.

It should be noted that this mechanism for transformation targeting is not
limited to matching substitutions rules. Similar to substitution rules,
instructions also have names and tags, and the same notation applies.
For example, a specific instruction ID can be matched directly as
`\verb|instruction_id|`, and all instructions whose tags match a given one
may be matched by `\verb|*$instruction_tag|`, where the wildcard \verb|*|
for the instruction ID does not impose any matching constraint.
\subsection{Substitution rules for precomputation}
\label{sec:precomp}
For computations that make use of the same intermediate results
multiple times, it may be
desirable to store these results in some form of
temporary memory until they are needed again. Similarly, computations
targeting cache-constrained architectures that reference the same
off-chip data repeatedly may want to allocate on-chip temporary memory
to avoid incurring the fetch latency for this data again and again.
This challenge is met by \loopy's \verb|precompute()| transformation,
which generally helps programs trade off increased on on-chip
storage against the cost off repeatedly fetching or computing needed
intermediate results.

To facilitate precise targeting of precomputation, \loopy's
\verb|precompute()| transformation operates exclusively on substitution
rules. Any subexpression for which precomputation is desired must first be
converted to a substitution rule using the machinery of
Section~\ref{sec:subst-rules}.

Once a substitution rule has been created, the \verb|precompute|
transformation can be used to allocate storage and create instructions to
store the precomputed values. This is straightforward if the substitution
rule simply represents a scalar value. More interesting cases arise if the
value of the rule or one of its invocation arguments involve inames.
In this case, a set of inames can be provided to \verb|precompute()| which,
when swept out, generate all values of the substitution rule which are to
be precomputed. In this situation, enough storage is allocated to
accommodate the access footprint, and an auxiliary set of inames is
generated that sweep out the footprint and drive the precomputation.
Naturally, the precomputation logic can be applied with the same
fine-grained targeting described in Section~\ref{sec:ttp}.
\section{Some examples}
\label{sec:examples}
\subsection{Forward differencing}
Consider this example program which computes forward differences on a
1-dimensional array of length \verb|n|:
\begin{tcolorbox}[listingbox]
\begin{lstlisting}[style=custompython,gobble=2]
  knl = lp.make_kernel(
      "{[i]: 0<=i<n}",
      "result[i] = u[i+1]-u[i]")
\end{lstlisting}
\end{tcolorbox}
\noindent
Since each entry of \verb|u| is used twice, a plausible optimization for
parallel
architectures with limited caches (such as GPUs) is to store a group of
values of \verb|u| in storage closer to the processor.

To achieve group-wise prefetching,
we split the iteration domain into fixed-size pieces of length 16,
assuming divisibility to ease understanding by
avoiding the generation of many conditionals:
\begin{tcolorbox}[listingbox]
\begin{lstlisting}[style=custompython,gobble=2]
  knl = lp.split_iname(knl, "i", 16)
  knl = lp.assume(knl, "n mod 16 = 0")
\end{lstlisting}
\end{tcolorbox}
\noindent
Next, we extract the access to \verb|u| into a substitution rule
\verb|u_acc| and apply (sequential, for now) precomputation for each sweep
through iterations of \verb|i_inner|, and assuming all other inames remain
constant:
\begin{tcolorbox}[listingbox]
\begin{lstlisting}[style=custompython,gobble=2]
  knl = lp.extract_subst(knl, "u_acc", "u[j]", parameters="j")
  knl = lp.precompute(knl, "u_acc", "i_inner", default_tag=None)
\end{lstlisting}
\end{tcolorbox}
\noindent
We obtain the following C code:
\begin{tcolorbox}[listingbox]
\begin{lstlisting}[style=customc,gobble=2]
  float u_acc_0[17];

  for (int i_outer = 0;
    i_outer <= -1 + int_floor_div_pos_b(15 + n, 16);
    ++i_outer)
  {
    for (int j = 0; j <= 16; ++j)
      u_acc_0[j] = u[j + 16 * i_outer];
    for (int i_inner = 0; i_inner <= 15; ++i_inner)
      result[i_inner + i_outer * 16] = 
        u_acc_0[1 + i_inner] + -1.0f * u_acc_0[i_inner];
  }
\end{lstlisting}
\end{tcolorbox}
Precomputation has found the (17-long) footprint of the access to \verb|u|
for each group of 16 iterations through \verb|i_inner|, created a suitable
prefetch loop, and modified the variable references to match.
Parallelization can the be applied to the generated loops as described
in \citep{kloeckner_loopy_2014}, as demonstrated in the following example.
\subsection{Matrix-Matrix multiplication}
This end-to-end Fortran-to-GPU example parallelizes a matrix-matrix
multiplication loop for parallel execution, matches the access to each of
the argument matrices into a substitution rule, and performs a block-wise
prefetch:
\begin{tcolorbox}[listingbox]
\begin{lstlisting}[style=customfortran,gobble=2]
  subroutine dgemm(m,n,l,alpha,a,b,c)
    implicit none
    real*8 temp, a(m,l),b(l,n),c(m,n), alpha
    integer m,n,k,i,j,l

    do j = 1,n
      do k = 1,l
        do i = 1,m
          c(i,j) = c(i,j) + alpha*b(k,j)*a(i,k)
        end do
      end do
    end do
  end subroutine

  !$loopy begin transform
  ! dgemm = lp.split_iname(dgemm, "i", 16,
  !    outer_tag="g.0", inner_tag="l.1")
  ! dgemm = lp.split_iname(dgemm, "j", 8,
  !    outer_tag="g.1", inner_tag="l.0")
  ! dgemm = lp.split_iname(dgemm, "k", 32)
  !
  ! dgemm = lp.extract_subst(dgemm,
  !    "a_acc", "a[i1,i2]", parameters="i1, i2")
  ! dgemm = lp.extract_subst(dgemm,
  !    "b_acc", "b[i1,i2]", parameters="i1, i2")
  ! dgemm = lp.precompute(dgemm,
  !    "a_acc", "k_inner,i_inner")
  ! dgemm = lp.precompute(dgemm,
  !    "b_acc", "j_inner,k_inner")
  !$loopy end transform
\end{lstlisting}
\end{tcolorbox}
\section{Conclusions}
\label{sec:conclusions}
\loopy\ provides a small, composable code generation capability for
high-performance array code on CPU- and GPU-type shared memory
parallel computers. It is available under the MIT open-source
license from \url{http://mathema.tician.de/software/loopy}.

The core contributions described in this article and implemented in
\loopy\ are the following:
\begin{itemize}
  \item An transformation targeting scheme based on substitution rules and
  `tags' that can be used to very precisely specify what parts of an
  expression in a program is to be transformed.
  \item A way of using a high-level language (Python in this
  instance) in a `pragma'-like role, for the transformation of a
  program in a lower-level kernel language.
  \item A translation scheme from a subset of Fortran into \loopy's
  polyhedral-like representation.
  \item The use and extraction of substitution rules to capture precisely
  what elements of a computation should be precomputed, and across
  which dependent axes.
\end{itemize}

\loopy's kernel representation, its library of transformations, and its
runtime features combine to provide a compelling environment within which
array-shaped computations can be conveniently expressed and optimized.
Some examples in \citep{kloeckner_loopy_2014} illustrated that
high-performance variants are within the set of programs reachable via
\loopy\ transformations. This article described some techniques to
help broaden the set of codes that can benefit from these transformations,
providing a pathway to performance that does not compromise
maintainability and separation of concerns.


\acks%

I would like to acknowledge tremendously influential discussions with Tim
Warburton that led to the genesis of \loopy\ and guided its design. I would
also like to acknowledge feedback from early adopters of \loopy, including
Rob Kirby, Lucas Wilcox, Maxim Kuznetsov, and Ivan Oseledets.

My work on \loopy\ was supported in part by US Navy ONR grant number
N00014-14-1-0117, and by the National Science Foundation under grant number
DMS-1418961.

\bibliographystyle{abbrvnat}


\bibliography{loopy}

\end{document}